# Ab initio geometry and bright excitation of carotenoids: Quantum Monte Carlo and Many Body Green's Function Theory calculations on peridinin

*Emanuele Coccia[1], Daniele Varsano[2] and Leonardo Guidoni[1]*

[1]Dipartimento di Scienze Fisiche e Chimiche, Università degli Studi dell'Aquila, via Vetoio, 67110 L'Aquila, Italy

[2]S3 Center, CNR Institute of Nanoscience, Via Campi 213/A, 41125 Modena, Italy

ABSTRACT

In this letter we report the singlet ground state structure of the full carotenoid peridinin by means of variational Monte Carlo (VMC) calculations. The VMC relaxed geometry has an average bond length alternation of 0.1165(10) Å, larger than the values obtained by DFT (PBE, B3LYP and CAM-B3LYP) and shorter than that calculated at the Hartree-Fock (HF) level. TDDFT and EOM-CCSD calculations on a reduced peridinin model confirm the HOMO-LUMO major contribution of the $B_u^+$-like ($S_2$) bright excited state. Many Body Green's Function Theory (MBGFT) calculations of the vertical excitation energy of the $B_u^+$-like state for the VMC structure (VMC/MBGFT) provide excitation energy of 2.62 eV, in agreement with experimental results in *n*-hexane (2.72 eV). The dependence of the excitation energy on the bond length alternation in the MBGFT and TDDFT calculations with different functionals is discussed.

Carotenoids are among the most abundant chemical species in biological systems. Similarly to chlorophylls, they play a fundamental role in light harvesting and energy transfer mechanisms in photosynthetic organisms.[1] In addition, carotenoids get the necessary photoprotective activity of quenching triplet chlorophylls and singlet oxygen molecules.[2] Although carotenoids are characterized by a large variety in terms of conjugation length of the polyenic chain and of substituents, they share some properties like the state ordering and the character of the singlet and triplet excitations. In the present letter we consider the carotenoid peridinin (**PID,** Chart 1), present in the Peridinin-Chlorophyll-a protein (PCP), a water-soluble complex deriving from marine dinoflagellate *Amphidinium carterae* containing the highest peridinin to chlorophyll-a ratio in nature, namely 4:1 for each domain.[3] After a single-photon singlet excitation $S_0 \rightarrow S_2$ ($B_u^+$-like, one-photon allowed), the chromophores couple with an adjacent chlorophyll-a according to a resonant energy transfer mechanism. Two energy transfer channels have been experimentally detected:[2-7] the first, accounting for about the 25% of energy[6] is the direct transfer from $S_2$ to the $Q_x$ state of the chlorophyll; the second very efficient route starts from a fast internal conversion from $S_2$ to $S_1$ ($A_g^-$-like, one-photon forbidden), eventually interacting with an intramolecular charge transfer (ICT) state and coupling with $Q_y$ of the chlorophyll.[8-13] Measurements in several solvents have been reported in the literature:[14] values for $S_1$ and $S_2$ states in *n*-hexane (2.0-2.3 and 2.56 eV,[1a,14] respectively) correspond to the spectral origin, the 0-

0 transition, while the vertical Franck-Condon maximum for $S_2$ is given by the peak at 2.72 eV.[1a,2,10]

Concerning the nature of the low-lying singlet excitations, the $S_1$ state of **PID** is characterized by covalent configurations of double excitation character (HOMO$^2$->LUMO$^2$) which lower the excitation energy thanks to a favorable mixing with single-electron excitations of HOMO-1->LUMO and HOMO->LUMO+1 character,[11,15-17] and can be directly investigated by two-photon spectroscopic techniques.[18,19] HOMO->LUMO transition instead dominates the $S_2$ state.[11,17] The dark $A_g^-$ state of polyenes with conjugation length $N>4$ lies in energy below the bright $B_u^+$ state, thanks to a large singlet-triplet splitting:[16(a)] for N=4 the two states are near-degenerate at CASPT2 and CC3 level,[15] whereas TDDFT[20] calculations show an inverted ordering with the dark $A_g^-$ state higher in energy, independently of the chosen functional.[15] The same occurs for the energies of $B_u^+$-like and $A_g^-$-like states in carotenoids, **PID** included. The same problem, e. g. the difficulty of TDDFT in predicting the correct level ordering, is also found in the characterization of $^1L_a$ and $^1L_b$ excited states of nonlinear acenes.[16b]

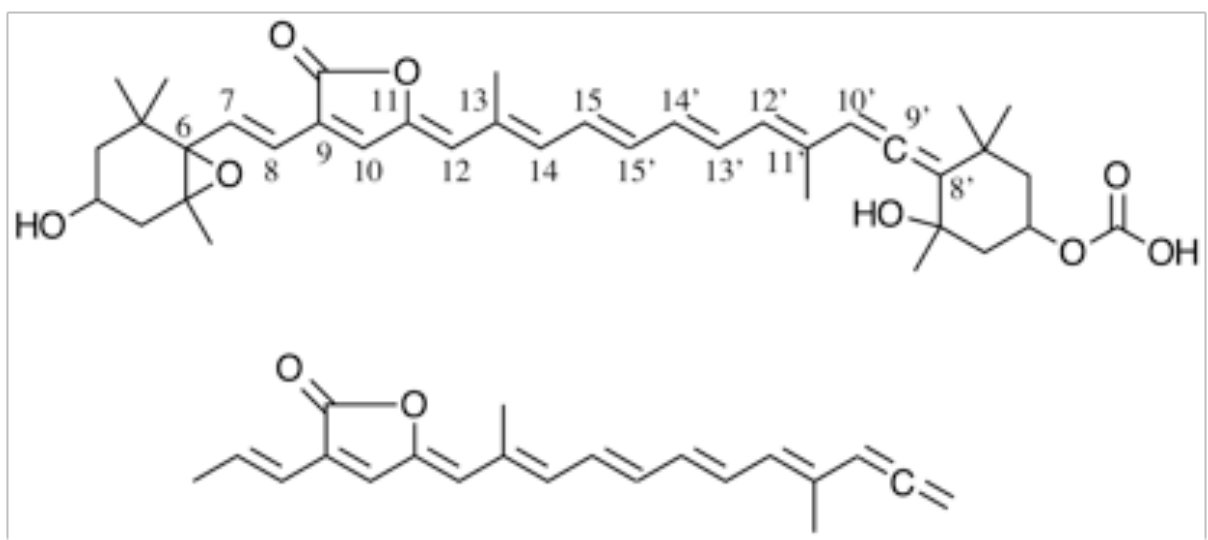

**Chart 1.** Sketch representation of peridinin (**PID**) and of the reduced peridinin model (**PID1**).

The accurate determination of the relaxed ground state geometry of carotenoids still represents a challenge for quantum chemistry calculations, due to the difficulties in the correct description of electronic correlation of the conjugated polyenic chain. Nevertheless, structural effects play an important role on the spectral tuning of carotenoids in gas phase, in solution, or in their protein environment and a reference high-level structure would be desirable. A key structural parameter in the spectral tuning of linear chromophores is the average bond length alternation (BLA), defined as the difference between the average of single bond and double bond distances, excluding (in the case of PID) the terminal double bond of the allene group. Differences of few hundredths of Å in average BLA may significantly alter vertical excitation energies of **PID**, since the molecular orbitals involved in the low-lying region of the absorption spectrum are delocalized along the polyenic chain. Geometries based on DFT calculations strongly depend on the choice of the functional: GGA and B3LYP functionals tend to overestimate the electronic delocalization on the carbon chain,[21,22] whereas CAM-B3LYP functional seems to give results in good agreement with experiments for linear all-*trans* polyenes.[22] So far accurate high-level quantum chemistry calculations (multireference or Coupled Cluster methods) are missing, due to their prohibitive computational cost on such large system size. Thanks to its capability in exploiting the High Performance Computing facilities based on massively parallel PetaFlop machines, Monte Carlo methods[23] can be considered a valid and fully *ab initio* alternative for quantum chemical calculations. We use here the Variational Monte Carlo (VMC) approach[23,24] to optimize the full **PID** ground state geometry by the Jastrow Antisymmetrized Geminal Power[24] wave function that has been shown to provide accurate results for several molecular properties.[25,26]

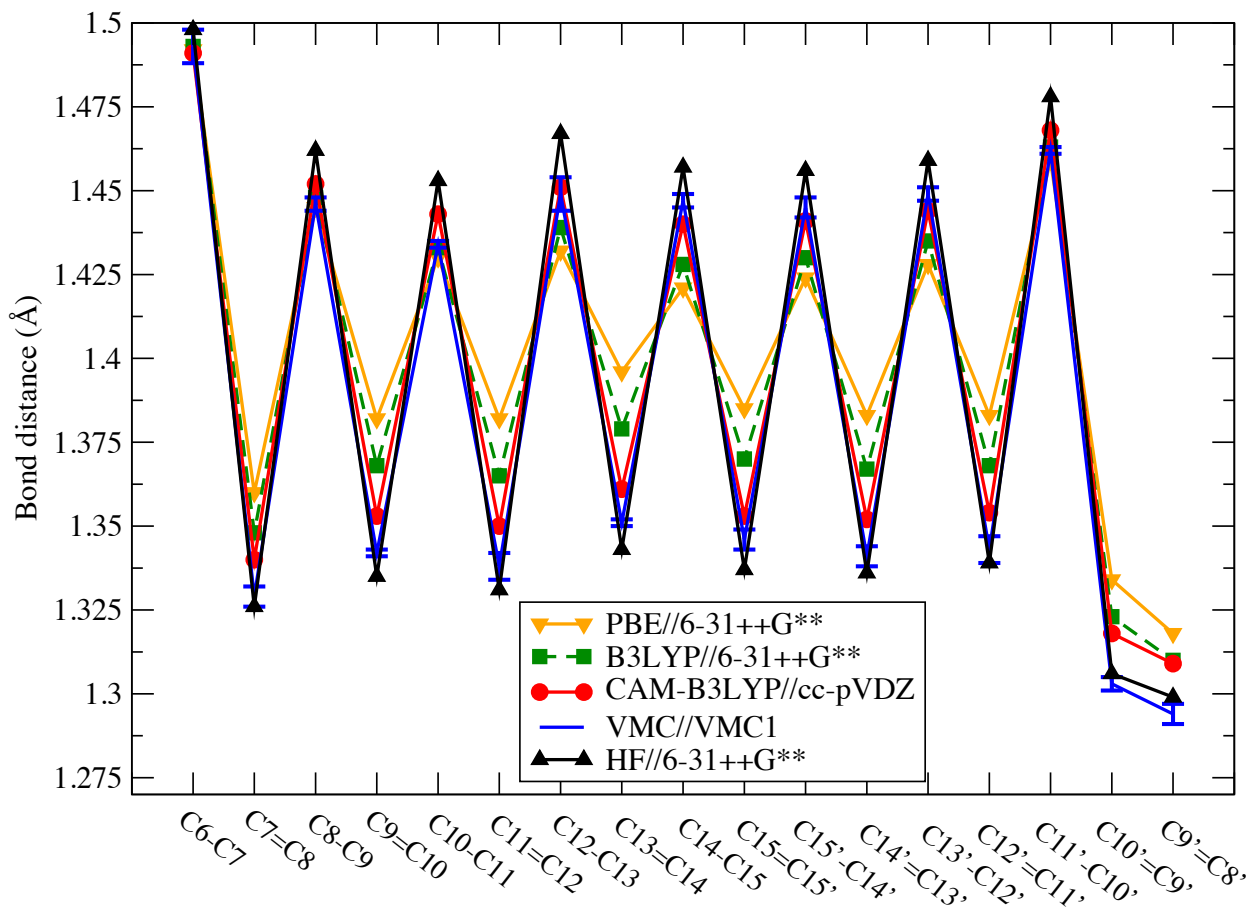

**Figure 1.** Comparison of the bond length alternation pattern of **PID** from ground state structures optimized at DFT/PBE, DFT/B3LYP, DFT/CAM-B3LYP, VMC and Hartree-Fock level. VMC1 basis is defined in Table 1 of SI and in refs 25c and 26.

Figure 1 reports the bond lengths of the optimized geometry of the VMC structure (blue in the color version) as well as a comparison with structures obtained at DFT (PBE//6-31++G**, B3LYP//6-31++G** and CAM-B3LYP//cc-pVDZ) and Hartree-Fock (HF//6-31++G**) levels. In terms of average BLA, the VMC value is 0.1165(10) Å, as expected smaller than the HF limit (0.135 Å) and higher than the PBE (0.066 Å), B3LYP (0.085 Å) and CAM-B3LYP (0.106 Å) values. Whereas the PBE functional produces the smallest difference between single and double bonds, as well known from theoretical investigations on polyacetilenes,[21b] the geometrical parameters of the CAM-B3LYP are quite similar to those obtained by VMC, as already observed in the gas phase structure of the retinal protonated Schiff Base,[26] even though VMC double bonds are seen to be shorter.

**Table 1.** EOM-CCSD excited states energies (eV) of **PID1**, calculated using a 16+16 orbital window, as a function of the basis sets. Oscillator strengths *(f)* in italics.

| Geometry | B3LYP | VMC |
|---|---|---|
| 3-21G | 3.84, *0.06* <br> 4.41, *2.99* | 4.42, *0.36* <br> 4.80, *2.82* |
| 6-31G | 3.82, *0.09* <br> 4.32, *3.04* | 4.38, *0.55* <br> 4.72, *2.66* |
| 6-311G | 3.81, *2.39* <br> 4.08, *1.68* | 4.20, *3.26* <br> 4.69, *0.82* |
| D95 | 3.84, *0.22* <br> 4.20, *3.08* | 4.37, *1.31* <br> 4.65, *2.06* |
| cc-pVDZ | 3.87, *2.53* <br> 4.11, *1.34* | 4.26, *3.34* <br> 4.72, *0.54* |
| cc-pVTZ | 3.81, *3.51* <br> 4.18, *0.63* | 4.19, *3.72* <br> 4.83, *0.39* |

The reduced model (**PID1** in Chart 1) is fully representative of the low-lying electronic excitations of **PID**, as already shown by Wagner et al.[13] Despite standard TDDFT is known to fail in the prediction of the energy inversion between $S_1$ and $S_2$ states,[15] it can provide a satisfactory description of the bright excitation. On the other hand, the well behaved performance of TDDFT at LDA level using the Tamm-Dancoff approximation (TDA) for **PID**[11] has been demonstrated to derive from a fortuitous cancellation of errors.[16a] TDDFT//6-31++G** calculations on **PID1** for the singlet excitations on the five structures give evidence (Tables 7 and 8 in Supporting Information (SI)) of a dominant HOMO->LUMO contribution in the $B_u^+$-like state, lower in energy than the $A_g^-$-like state, independently on the chosen functional and geometry.

Semi-empirical MNDO/PSDCI[10,13,27,28] and EOM-CCSD[29] methods show a good agreement with experimental findings in terms of energy values and intensity of the transitions:[10,13] using the D95 basis on a B3LYP geometry, the EOM-CCSD $B_u^+$-like excitation energy has been found to be around 2.5 eV with the $A_g^-$-like energy close to 2.2 eV, if the MP2 ground state is used as reference. However, Krylov asserts that doubly excited dark states in polyenes ($A_g^-$) are poorly described by EOM-CCSD since the important configurations appear at different level of excitation, providing a not balanced description of the excitation.[29d]

Basis sets and size of the active space can severely affect the convergence of EOM-CCSD excitation energies. Use of the **PID1** model was needed in order to reduce the computational effort. An orbital window of 16 highest occupied orbitals and 16 lowest virtual orbitals has been selected, following the procedure of ref 13. A convergence study of the first two excited states with respect to the basis set is reported in Table 1; similar analysis for a 8+8 and a 32+32 orbital window can be found in Tables 10 and 11 in SI. At variance with the analysis in ref 13, the excitation energies are consistently referring to EOM-CCSD ground state energy and not to the MP2 ground state energy.

Calculations on the B3LYP and VMC structures confirm within the EOM-CCSD framework that the $B_u^+$-like bright excited state has a dominant HOMO-LUMO contribution, whereas the $A_g^-$-like state, together with HOMO-1->LUMO and HOMO->LUMO+1 transitions, has a significant HOMO$^2$->LUMO$^2$ double-excitation character. All the excitation energies are anyway systematically overestimated and the use of larger basis sets, like cc-pVDZ and cc-pVTZ, results in an incorrect level ordering, with $B_u^+$-like state lower in energy than the $A_g^-$-like one. When the excitation operators are applied to a larger (32+32) orbital window (Table 11 in SI), EOM-CCSD calculations find the right level ordering for all the considered basis sets for the B3LYP geometry, even if the energies remain too high. The same effect is not observed for the VMC structure, evidence that the geometrical parameters play a fundamental role in the theoretical characterization of low-lying singlet states of peridinin and carotenoids. The failure in finding a robust and clear convergence as a function of the basis set complexity may be ascribed to the small orbital space, to the unbalanced description of configurations with the same excitations within the EOM-CCSD scheme and to the use of a mean-field (HF) reference. In the light of our results, the agreement between the calculations and the experimental excitation reported in ref 13 might be due to a combination of several aspects: the size of the orbital window, the choice of the basis set, the BLA pattern of the ground state geometry, and the fact that the reported excitations are calculated as energy differences between EOM-CCSD excited state and MP2 ground state energies.

**Table 2.** TDDFT excited states energies (eV) of **PID**, calculated using SVWN, B3LYP, CAM-B3LYP functionals. 6-31++G** basis set has been used for all the calculations. Oscillator strengths *(f)* in italics.

| Geometry | PBE | B3LYP | CAM-B3LYP | VMC | HF |
|---|---|---|---|---|---|
| **TD-SVWN** | 2.02, *2.95* <br> 2.22, *0.39* | 2.08, *2.61* <br> 2.34, *0.61* | 2.13, *2.20* <br> 2.47, *0.81* | 2.18, *2.08* <br> 2.55, *0.78* | 2.23, *1.85* <br> 2.65, *0.91* |
| **TD-B3LYP** | 2.20, *3.45* <br> 2.76, *0.20* | 2.31, *3.29* <br> 2.91, *0.31* | 2.44, *3.03* <br> 3.09, *0.48* | 2.52, *2.95* <br> 3.19, *0.52* | 2.63, *2.79* <br> 3.32, *0.62* |
| **TD-CAM-B3LYP** | 2.42, *3.66* <br> 3.65, *0.09* | 2.61, *3.62* <br> 3.86, *0.09* | 2.82, *3.55* <br> 4.07, *0.02* | 2.93, *3.53* <br> 4.18, *0.00* | 3.10, *3.51* <br> 4.30, *0.03* |

Moving to the excited states of the full chromophore, Table 2 summarizes the energy and oscillator strengths of the first two electronic excitations investigated at the TDDFT//6-31++G** level, using Gaussian 09 package[30] employing three different functionals (SVWN, B3LYP, and CAM-B3LYP) on five different geometries (PBE, B3LYP, CAM-B3LYP, VMC and HF). TDDFT calculations allowed us to investigate the performance of several functionals (LDA, GGA, hybrid and long-range corrected hybrids) on both geometry and excited states of **PID**.

A systematic blue shift in the excitation energies is observed when increasing the average BLA, i.e. moving from PBE to HF geometry (Figure 2 and Table 2), as also described by a very recent DFT/MRCI investigation on **PID**.[31] In all cases the order of states appears wrong, the $A_g^-$-like being, independently of the specific combination of functional and geometry, too high in energy, as already found for the reduced model. Excitations, as expected, are also strongly dependent on the chosen functional. Looking at the $B_u^+$-like state (the lowest in energy in our calculations) in Table 2, the SVWN functional tends to underestimate the vertical energy by ≈0.5-0.7 eV. TDDFT/B3LYP and TDDFT/CAM-B3LYP on the VMC structure (2.52 and 2.93 eV) differ from the experimental 2.72 eV by ≈0.2 eV, respectively underestimating and overestimating the excitation energy. None of the TDDFT vertical energies does lead to a satisfactory agreement when the VMC structure is used. Since the choice of the functional influences both the BLA and the excited state energy, different combinations of functionals for the ground state geometry optimization and for the TDDFT calculations can match the experimental value, although the agreement has to be considered fortuitous, as in the case of the CAM-B3LYP excitation on the B3LYP geometry (2.61 eV) and the B3LYP excitation on HF geometry (2.63 eV). On the other hand, the bright excitation energy obtained by using the CAM-B3LYP functional for both geometry optimization and excited states calculation (2.82 eV) is comparable to the experimental absorption, with a discrepancy of 0.1 eV. In spite of the good performance of CAM-B3LYP for the excited state calculations of **PID**, the same functional seems to overestimate the $S_1$ state energy of retinal and other positively charged retinal and GFP-like models, as reported in references.[26,32]

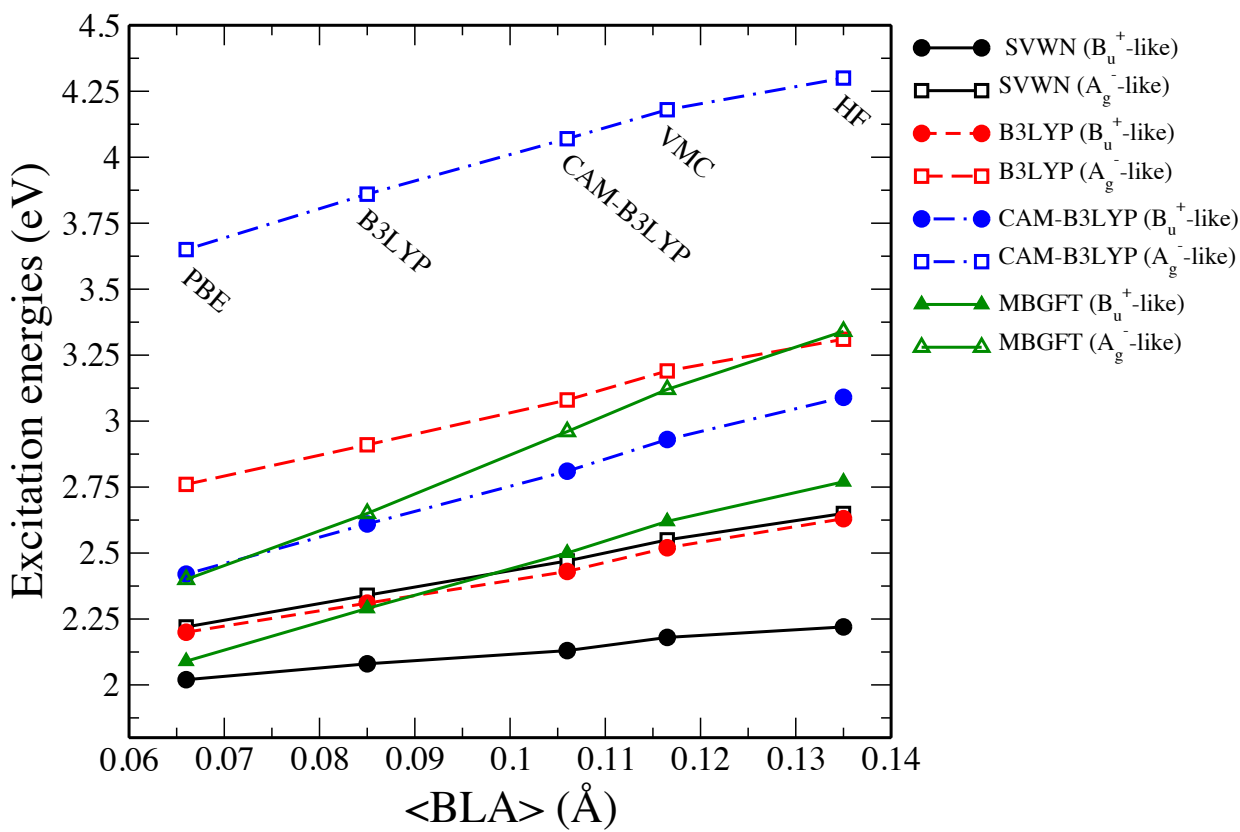

**Figure 2.** $B_u^+$-like and $A_g^-$-like excited state energies (eV) of **PID** calculated by TDDFT//6-31++G** with three different functionals (SVWN, B3LYP and CAM-B3LYP) as a function of the average BLA (Å). MBGFT results are also shown.

It is interesting to note that the oscillator strength of $B_u^+$-like ($A_g^-$-like) slightly decreases (increases) for the VMC and HF structures with respect to the DFT geometries; no switch between the two states is however observed. Table 9 in Supporting Information (SI) reports the main components of the excitations in terms of molecular orbitals: i) the $B_u^+$-like state is dominated by a HOMO->LUMO transition; ii) HOMO-1->LUMO and HOMO->LUMO+1 components characterize the $A_g^-$-like; iii) structural changes do affect the transition intensity but not the character of the first two states.

TD-B3LYP estimation of the adiabatic transition to the $S_2$ state performed on the B3LYP structures corresponds to 2.21 eV, 0.10 eV smaller than the vertical one (2.31 eV, Table 2), fully consistent with the shift of 0.16 eV experimentally observed.[2]

**Table 3.** Kohn-Sham gap (LDA), $G_0W_0$ gap and excited states energies (for the first two states) by solving BS equation for **PID**. All the energies are in eV.

| Geometry | PBE | B3LYP | CAM-B3LYP | VMC | HF |
|---|---|---|---|---|---|
| **Kohn-Sham Gap** | 1.20 | 1.34 | 1.49 | 1.56 | 1.69 |
| **$G_0W_0$ Gap** | 4.16 | 4.41 | 4.71 | 4.89 | 5.11 |
| **Excited states (BS)** | 2.09 | 2.29 | 2.50 | 2.62 | 2.77 |
|  | 2.40 | 2.65 | 2.96 | 3.12 | 3.34 |

Since the application of several combinations of functionals (for TDDFT) and ground state structures gives rise to a wide range of values for the bright excitation energy, a step further in the characterization of the $B_u^+$-like state is possible by means of the use of the Many Body Green's Function Theory (MBGFT) methods for excited states.[33] Such techniques have been successfully used to describe with high accuracy quasiparticle energies and optical excitations in several materials,[33] including polyenic chains where TDDFT with local and semi-local approximation dramatically fails.[34] Due to the high computational cost, very few calculations are present in the literature on biomolecules[35] showing a remarkable agreement with experiments. The MBGFT excitation energies have been calculated using the plane waves Yambo code.[36] The absorption spectra have been calculated considering a Kohn-Sham system within the Local Density Approximation (LDA) as a zero order non-interacting Hamiltonian using the Quantum Espresso package.[37] The quasiparticle corrections have been evaluated within the $G_0W_0$ approximation for the self-energy operator and the electron-hole screened interaction is included by solving the Bethe-Salpeter (BS) equation, where the static screening is calculated within the random-phase approximation. As observed in ref 35 for other biomolecules we have also found that TDA is not suitable for the **PID** system. Details on the implementation of the GW/BS calculations can be found in ref 36, and the computational details are reported in the SI. In Table 3 electronic gaps at the Kohn-Sham level and $G_0W_0$ corrected are reported for all the studied structures, together with the excitation energies calculated at BS level. Quasiparticle energies of the frontier orbitals are shown in Figure 1 of SI. We observe that the electronic gap, already at DFT level, increases with the increasing of the BLA and the effect is enhanced when quasiparticle corrections are included. The blue shift in the excitation energies is also observed at BS level going through increasing BLA values. Noticeably the dependence of the $B_u^+$-like energy on the average BLA calculated by MBGFT methods is enhanced with respect to TDDFT calculations indicating an even larger role of structural effects in the tuning of absorption properties of **PID**.

The BS absorption spectra are collected in Figure 2 of SI. The bright excitation of 2.62 eV on the VMC structure is mainly due to the HOMO->LUMO transition (88%), showing a good agreement with the $B_u^+$-like vertical experimental energy (2.72 eV) measured in *n*-hexane. The latter value can be taken as a reference for gas phase calculations of the vertical excitation to the bright state, assuming small electrostatic coupling between *n*-hexane and **PID** and zero point energy contributions. Nevertheless, it is worth to mention that vibronic terms could have an appreciable weight in the applied theoretical model,[38] and a deeper and more quantitative insight on these effects requires a further investigation. In summary, the use of high-level geometry optimization such as VMC in combination with a converged treatment by first principles of electronic excitations by MBGFT is able to provide an accurate estimation of the bright electronic excitation in **PID**. In order to correctly describe the double excitation character of the $A_g^-$-like state in MBGFT framework the use of a dynamic kernel in the BS equation is required.[39]

In this Letter we have reported the ground state optimized geometry of a carotenoid at the Quantum Monte Carlo level of theory, which is becoming an affordable many-body technique for the study of large molecular systems, where electron correlation plays an important role. The combined use of the VMC structure and MBGFT methods (VMC/MBGFT) represents a fully *ab initio* approach to obtain excitation energies in meaningful agreement with the experimental findings, overcoming the difficulties in the choice of the proper functional in the TDDFT framework and avoiding inclusion of parameters, as in the case of the hybrid CAM-B3LYP. This work can be considered as a needed step for the quantitative investigation of the effects of the protein environment of peridinin molecules and other carotenoids in photosynthetic complexes.

ASSOCIATED CONTENT

**Supporting Information**. VMC wave function, details on MBGFT calculations, bond distances for DFT, VMC and HF structures, transition components for TDDFT calculations on **PID** and **PID1**, MBGFT quasiparticle energies and absorption spectra, convergence tables as a function of basis set and active space for EOM-CCSD calculations on **PID1**. This material is available free of charge via the Internet at http://pubs.acs.org

AUTHOR INFORMATION

**Corresponding Author**

*leonardo.guidoni@univaq.it

**Author Contributions**

The manuscript was written through contributions of all authors. All authors have given approval to the final version of the manuscript.

**Funding Sources**

The European Research Council Project MultiscaleChemBio (n. 240624) within the VII Framework Program of the European Union has supported this work.

ACKNOWLEDGMENT

The authors acknowledge Prof. Sandro Sorella for the assistance in the use of the TurboRVB Quantum Monte Carlo code. DV acknowledge also support from *Futuro in Ricerca* grant No. RBFR12SW0J of the Italian Ministry of Education, University and Research, and European Community through the CRONOS project, Grant Agreement No. 280879. Computational resources were provided by the PRACE Consortium (Project PRA053), CINECA Computing Center and the Caliban-HPC Lab of the University of L'Aquila.